\documentclass[conference]{IEEEtran}

\usepackage{cite}

\usepackage{adjustbox}
\usepackage{balance}       
\usepackage{float}
\usepackage{gensymb}
\usepackage{tabularx}
\usepackage{svg}
\usepackage{algorithm}
\usepackage{amsmath}
\usepackage{threeparttable}
\usepackage{multirow}
\usepackage{booktabs}
\usepackage{algpseudocode}
\usepackage{makecell}
\usepackage{color, colortbl}
\usepackage{subfigure}
\usepackage[normalem]{ulem}
\usepackage{tikz, listofitems}
\usetikzlibrary{shapes.geometric, arrows}
\usepackage{caption}
\usepackage{subcaption}
\usepackage{graphicx}
\usepackage{ulem}
\usepackage{hyperref}
\usepackage{pifont}  
\usepackage{graphics}      

\usetikzlibrary{patterns} 
\usetikzlibrary[patterns] 
\usetikzlibrary{decorations.pathreplacing, calligraphy}
\usetikzlibrary {arrows.meta}

\definecolor{lightgray}{gray}{0.9}

\algnewcommand{\Inputs}[1]{%
  \State \textbf{Inputs:}
  \Statex \hspace*{\algorithmicindent}\parbox[t]{.8\linewidth}{\raggedright #1}
}
\algnewcommand{\Initialize}[1]{%
  \State \textbf{Initialize:}
  \Statex \hspace*{\algorithmicindent}\parbox[t]{.8\linewidth}{\raggedright #1}
}

\newcommand{\name}{\texttt{$\mu$Touch}}
\newcommand{\mdname}{\texttt{MagDelta}}

\definecolor{blue1}{rgb}{0.215, 0.683, 0.844}
\definecolor{airforceblue_darker}{rgb}{0.828,0.914,0.910}
\definecolor{airforceblue}{HTML}{D5F3F1}
\newcolumntype{g}{>{\columncolor{airforceblue}}c}
\newcolumntype{o}{>{\columncolor{airforceblue_darker}}c}
\newcolumntype{P}[1]{>{\raggedright\arraybackslash}p{#1}}  
\newcolumntype{L}[1]{>{\raggedright\arraybackslash}p{#1}}
\newcolumntype{R}[1]{>{\raggedleft\arraybackslash}p{#1}}

\graphicspath{{./figures/}}
\begin{document}


\title{$\mu$Touch: Enabling Accurate, Lightweight Self-Touch Sensing with Passive Magnets}



\author{
\IEEEauthorblockN{
Siyuan~Wang\IEEEauthorrefmark{1},
Ke~Li\IEEEauthorrefmark{2},
Jingyuan~Huang\IEEEauthorrefmark{1},
Jike~Wang\IEEEauthorrefmark{1},
Cheng~Zhang\IEEEauthorrefmark{2},
Alanson~Sample\IEEEauthorrefmark{3},
Dongyao~Chen\IEEEauthorrefmark{1}
}
\IEEEauthorblockA{
\IEEEauthorrefmark{1}Shanghai Jiao Tong University, China \quad
\IEEEauthorrefmark{2}Cornell University, USA \quad
\IEEEauthorrefmark{3}University of Michigan, USA
}
\IEEEauthorblockA{\footnotesize Email: \{wsy0227,aquamarine\_indigo,jikewang,chendy\}@sjtu.edu.cn\\
\footnotesize \hspace{3.1em} \{kl975,chengzhang\}@cornell.edu, apsample@umich.edu}
}







\maketitle

\begin{abstract}
Self-touch gestures (e.g., nuanced facial touches and subtle finger scratches) provide rich insights into human behaviors, from hygiene practices to health monitoring. However, existing approaches fall short in detecting such micro gestures due to their diverse movement patterns.  

This paper presents {\name}, a novel magnetic sensing platform for self-touch gesture recognition. {\name} features (1) a compact hardware design with low-power magnetometers and magnetic silicon, (2) a lightweight semi-supervised framework requiring minimal user data, and (3) an ambient field detection module to mitigate environmental interference.
We evaluated {\name} in two representative applications in user studies with 11 and 12 participants. 
$\mu$Touch only requires three-second fine-tuning data for each gesture — new users need less than one minute before starting to use the system. 
$\mu$Touch can distinguish eight different face-touching behaviors with an average accuracy of \textbf{93.41\%}, and reliably detect body-scratch behaviors with an average accuracy of \textbf{94.63\%}. 
{\name} demonstrates accurate and robust sensing performance even after a month, showcasing its potential as a practical tool for hygiene monitoring and dermatological health applications. Code is available at \url{https://wangmerlyn.github.io/muTouch/}.

\end{abstract}



\section{Introduction}


\label{sec:introduction}

Micro self-touch gestures (e.g., face touching or body scratching) are subtle yet informative behaviors that carry significant implications for personal hygiene and healthcare~\cite{Pang2022SelfTouch,HARRIGAN19851161}. 
According to the World Health Organization (WHO) prevention guidelines, minimizing face touching --- especially unconscious actions (e.g., nose rubbing, eye touching, or mouth stroking) --- is critical for controlling the spread of infectious diseases~\cite{who2020, kwok2015face}. 
Similarly, scratching or skin rubbing can also provide valuable insights into an individual’s physical and mental health. 
Uncontrolled scratching may interfere with wound healing and promote scarring~\cite{Gurtner2022}, aggravate dermatological inflammation~\cite{Zhang2024}, and is closely linked to psychological stress and anxiety through the itch–scratch cycle~\cite{Sanders2018, Dalgard2020}. 
Therefore, developing a practical method for self-touch gesture recognition holds profound promise for advancing personal hygiene monitoring and healthcare applications.

\begin{figure}[t]
    \centering
    \includegraphics[width=\linewidth]{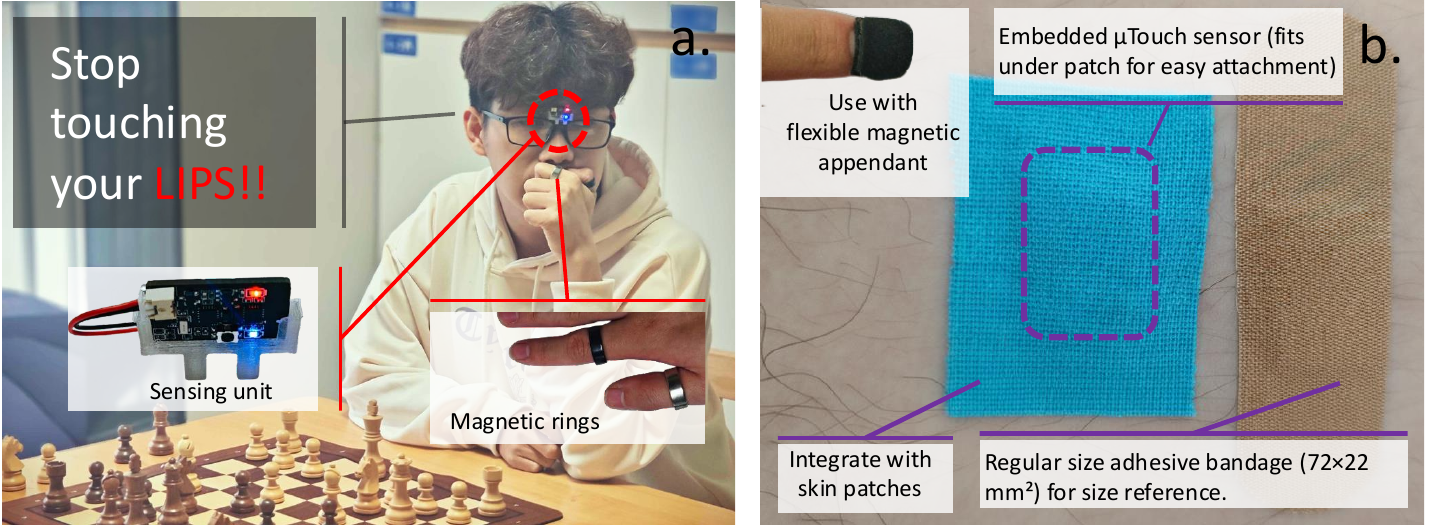}
\caption{Overview of {\name}. (a) Face-touching detection for infection prevention and (b) body-scratch monitoring for dermatological health. {\name} provides a unified magnetic sensing framework for accurate, privacy-preserving detection of self-touch behaviors.}
    \label{fig:user-overview}
    \vspace{-1em}
\end{figure}

However, recognizing self-touch gestures in real-world settings remains challenging due to the limitations of existing approaches.  
Vision- and radar-based methods, while expressive, are highly sensitive to environmental factors such as poor lighting or occlusion and often raise privacy concerns~\cite{Li2022NailRing:, Pittaluga2016Sensor-level, Pittaluga2017Pre-Capture, Ohn-Bar2014Hand, Choi2019Short-Range, oudah2020hand, kim2016hand}.  
Wearable motion sensors like IMUs overcomes occlusion but suffer from drift due to error accumulation during the integration of inertial sensor readings, undermining long-term reliability~\cite{Zhao2012Motion, Vaccaro2018Statistical, Cox2023Motion, zhang2021fine}.  
In addition, most systems rely on classifiers trained on large-scale generic datasets~\cite{D_touch, face_sense}, making efficient personalization difficult—particularly for face-touching and scratching behaviors that vary widely across individuals (e.g., which finger touches, frequency, intensity)~\cite{ozer2020Gesture}.  
As a result, a comprehensive per-user data collection is often required, which limits usability in health-related applications.  
Moreover, existing systems are typically tailored to specific body parts (e.g., the facial region), and their sensing pipelines and/or hardware designs are hard to generalize to other self-touch behaviors.  
They also fail under clothing coverage, which is unavoidable in many body self-touch scenarios.  
A summary of representative systems is shown in Table~\ref{tab:related}, highlighting that most lack one or more key properties.   

To overcome these limitations, we propose {\name}, a micro hand gesture recognition approach combining magnetic sensing and lightweight personal adaptation. {\name} delivers reliable detection, supports efficient personalization with few samples, and minimizes data collection overhead.

\noindent\textbf{Magnetic sensing hardware.}
{\name} consists of a low-power, wearable magnetic sensing unit, and a low-cost passive magnetic attachment. 
{\name}'s sensing unit consists of three low-power Hall-effect magnetometers. 
As we will elaborate in Sec.~\ref{sec:technical_hardware}, we employed a simulation-guided computational design to determine that three sensors are sufficient to achieve robust accuracy while maintaining a compact form factor. 
This design enables our system to capture subtle magnetic field variations within a miniaturized sensing unit of only 2.4$\times$1.2 cm$^2$.
Thus, it can be attached to everyday objects such as glass frames and wireless earbuds.
We propose two novel passive magnetic setups, i.e., magnetic rings and flexible silicons, to support more diverse application scenarios. 

\noindent\textbf{Lightweight detection pipeline.}
{\name} employs a lightweight detection pipeline that first uses a training-free trigger module (Sec.~\ref{sec:technical_algorithm}) to identify potential gesture events, before activating the classification module.
To bring personalization to the classification module without incurring heavy data collection overheads, we propose a self-supervised learning approach to train a \emph{magnetic motion encoder}, which learns compact feature representations of self-touch gestures without requiring extensive labeled datasets.
Building on the pre-trained encoder, users can fine-tune a downstream classifier with only three samples per gesture (recorded in under 10 seconds), enabling rapid personalization with minimal effort.
This efficiency enables users to fine-tune a classifier for six gestures in under a minute, minimizing data labeling efforts and allowing rapid adaptation to each user's unique gesture vocabulary.

To demonstrate {\name}’s potential in healthcare-related scenarios, we focus on two representative applications of self-touch detection:
	1.	\emph{\textbf{infection prevention}}, by detecting and minimizing unconscious face-touching behaviors that increase the risk of disease transmission; and
	2.	\emph{\textbf{dermatological and behavioral health monitoring}}, by identifying scratching behaviors that may interfere with wound healing, aggravate skin conditions, or signal stress-related habits.
These tasks (Fig.~\ref{fig:user-overview}) exemplify the broader applicability of {\name} in recognizing fine-grained self-touch gestures.

We tested {\name}'s performance in a user study with 11 participants.
In the face-touching detection task, {\name} achieved an accuracy of 93.41\% using two magnetic rings.
For the body-scratch detection task, {\name} achieved an accuracy of 94.63\% with a flexible magnetic silicon.
In a follow-up study conducted \emph{one-month later} in a completely new environment, {\name} maintained robust performance with accuracies of 90.50\% for face-touching detection and 92.59\% for body-scratch detection.
These results highlight {\name}'s effectiveness and versatility across tasks and environments.    

\begin{table*}[t]
\centering
\captionsetup{font=footnotesize, labelfont=bf, textfont=normal, skip=2pt}
\caption{Comparison of face-touching detection systems using different modalities. 
Blue-green (\textcolor[HTML]{00C4B3}{Yes/Low}) indicates positive attributes, 
while purple (\textcolor[HTML]{5E3C99}{No/High}) indicates negative attributes. 
We additionally evaluate whether systems can be extended to other self-touch behaviors 
and whether they remain effective under clothing coverage.}
\begin{tabular}{llccccccc}
\toprule
\multicolumn{1}{l}{\textbf{Name}} & \multicolumn{1}{l}{\textbf{Modality}} 
& \multicolumn{1}{l}{\textbf{Fine-grained}} 
& \multicolumn{1}{l}{\textbf{Few-shot}} 
& \multicolumn{1}{l}{\textbf{Customizable}} 
& \multicolumn{1}{l}{\textbf{Energy}} 
& \multicolumn{1}{l}{\textbf{Privacy preservative}} 
& \multicolumn{1}{l}{\textbf{Extendable}} 
& \multicolumn{1}{l}{\textbf{Clothing robust}} \\ 
\midrule
Covid-away~\cite{covid-away}       & accelerometer, ... & {\color[HTML]{5E3C99} No} & {\color[HTML]{5E3C99} No} & {\color[HTML]{5E3C99} No} & {\color[HTML]{00C4B3} Low}  & {\color[HTML]{00C4B3} Yes} & {\color[HTML]{5E3C99} No} & {\color[HTML]{00C4B3} Yes} \\
Saving-face~\cite{rojas2021scalable} & acoustic          & {\color[HTML]{5E3C99} No} & {\color[HTML]{5E3C99} No} & {\color[HTML]{5E3C99} No} & {\color[HTML]{00C4B3} Low}  & {\color[HTML]{00C4B3} Yes} & {\color[HTML]{5E3C99} No} & {\color[HTML]{5E3C99} No} \\
Face-sense~\cite{face_sense}        & thermal, EMG      & {\color[HTML]{5E3C99} No} & {\color[HTML]{5E3C99} No} & {\color[HTML]{00C4B3} Yes} & {\color[HTML]{5E3C99} High} & {\color[HTML]{5E3C99} No}  & {\color[HTML]{5E3C99} No} & {\color[HTML]{5E3C99} No} \\
D-touch~\cite{D_touch}              & infrared camera   & {\color[HTML]{00C4B3} Yes}& {\color[HTML]{5E3C99} No} & {\color[HTML]{00C4B3} Yes} & {\color[HTML]{5E3C99} High} & {\color[HTML]{5E3C99} No}  & {\color[HTML]{5E3C99} No} & {\color[HTML]{5E3C99} No} \\
\textbf{{\name}}                    & magnetic sensing  & {\color[HTML]{00C4B3} Yes}& {\color[HTML]{00C4B3} Yes}& {\color[HTML]{00C4B3} Yes} & {\color[HTML]{00C4B3} Low}  & {\color[HTML]{00C4B3} Yes} & {\color[HTML]{00C4B3} Yes}& {\color[HTML]{00C4B3} Yes} \\
\bottomrule
\end{tabular}
\label{tab:related}
\end{table*}



\section{Motivation and Background}
\label{sec:background}
\subsection{Self-Touch Gestures}
Self-touch gestures refer to habitual hand movements toward one’s own body, such as touching the face or scratching the skin~\cite{Streeck1993Gesture,HARRIGAN19851161}.
These movements are usually subtle and unnoticeable. 
For example, touching one's face or scratching the arm due to skin irritation.
However, categorizing these gestures is a non-trivial task.
We first analyze the unique challenges of detecting self-touches via a pilot study.
Next, we motivate {\name}'s magnetic sensing modality.

\subsection{Pilot Study of Self-Touches}
\label{sec:pilot-study}
We first selected face-touching as the pilot case of self-touches, given its high frequency and well-documented link to hygiene and infection risk.  
We observed 10 volunteers in two naturalistic settings (office and gym), recording their behaviors without disclosing the study focus to avoid bias.  
This pilot study revealed unique patterns of self-touch gestures, providing important insights for the design of {\name}.  

\begin{figure}[t]
    \centering
    \includegraphics[width=0.95\linewidth]{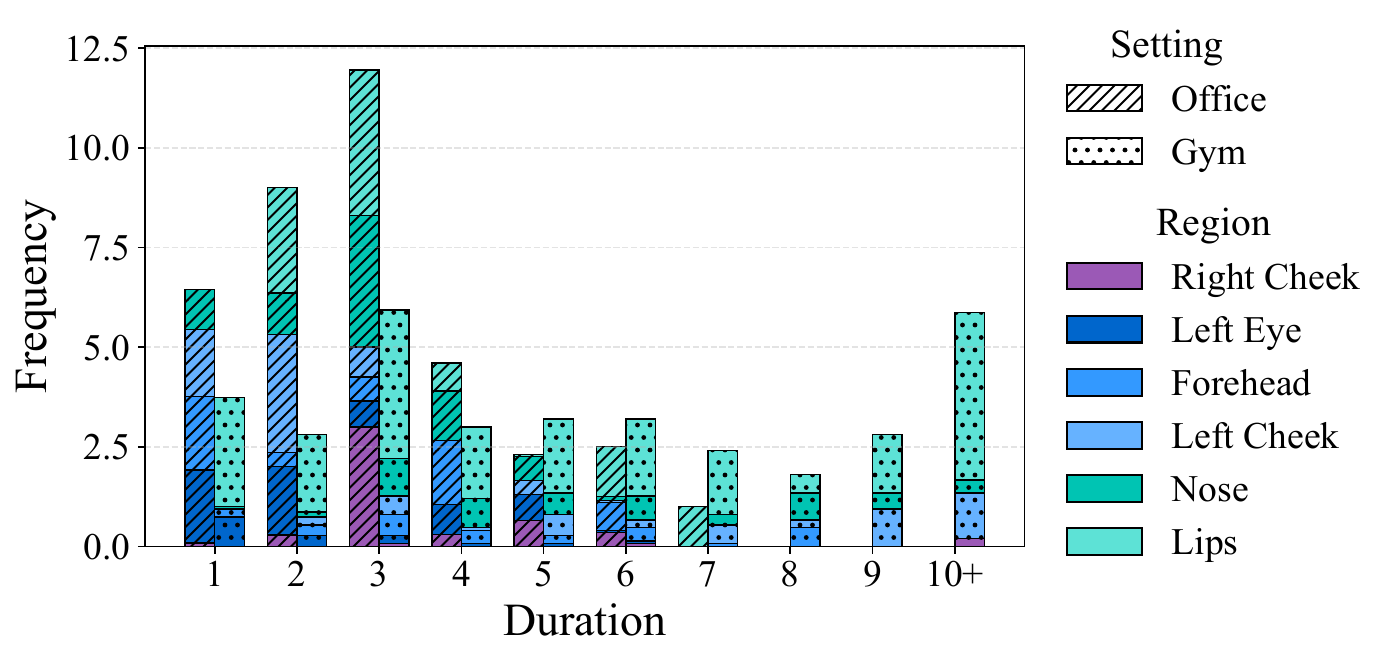}
    \caption{Pilot study of face-touching behaviors.}
    \label{fig:desk_and_gym}
\end{figure}

\noindent \textbf{Diversity.}
As shown in Fig.~\ref{fig:desk_and_gym}, the frequency and duration of different face-touching actions can have a wide range of diversity.
For instance, lip touching occurs on average 9.3 times per hour in the gym, with durations from one to six seconds, while in the office, it occurs about 21 times per hour, ranging from one to over 10 seconds.
This variance suggests inconsistent duration across actions, making behavior prediction challenging.

\noindent \textbf{One Handedness.}
Self-touches are predominantly one-handed, with users favoring their non-dominant hand more often. On average, users performed face touching 35 times with their non-dominant hand in an hour, compared to only 11 times with their dominant hand.
This finding echoes with existing studies~\cite{zhang2020most}.

In the following, we elaborate unique advantages of magnetic sensing for detecting micro gestures.

\subsection{Why Magnetic Sensing}
We employ the magnetic sensing approach for micro self-touch gesture detection for the following reasons:


\noindent\textbf{Reliability.}  
Magnetic field signals are highly sensitive to the distance between the sensor and magnet, which {\name} leverages to differentiate fine-grained gestures such as a gentle ear tap.  
Unlike visual systems, magnetic sensing is unaffected by occlusion or poor lighting, while also inherently safeguarding user privacy.
It further remains reliable under clothing coverage, which is crucial for body-related self-touch behaviors where garments would otherwise block visual or contact-based methods. 
A key challenge, however, is that the signal magnitude can be comparable to the ambient environmental field\cite{Golestani2021MItracking,app13095544,lyons2020iswc}.  
To address this, we design an environmental magnetic field mitigation algorithm (Sec.~\ref{subsec:tackling_the_environment_interference}) that effectively reduces ambient interference and enhances robustness.

\noindent\textbf{Low overheads.}
Camera-based methods like Face-sense~\cite{face_sense} consume significant power (e.g., 289 mW) as listed in Table~\ref{tab:related}.
In contrast, {\name} uses compact MEMS magnetic sensors with an average power consumption of 25 mW, ten times lower than vision-based approaches, making {\name} suitable for continuous use.


\section{Algorithm Design}
\label{sec:technical_algorithm}
The key challenge of {\name} is overcoming environmental noises caused by nearby electronic devices, power lines, and metal structures. 
In the following, we first elaborate {\name}'s design for overcoming interferences. Next, we present the machine-learning pipeline.
\subsection{Interference Mitigation}
\label{subsec:tackling_the_environment_interference}
Based on the magnetometer (e.g., MLX90393 by Melexis)'s datasheet, under the experimental conditions with a sampling rate set at 17 Hz, the sensor readings exhibit variances on each axis, quantified as (0.6, 0.6, 1.1) $\mu$T, indicating the level of noise present in the measurements.
Note that environmental and sensor noises are the main components of the \emph{noises} in the sensor reading, which can be reduced by averaging multiple measurements or filtering~\cite{finexus, mag_x, auraring}. 



To this end, we decompose the measured signal $R_t$ into three components:
\begin{equation}
    R_t = B_t + B_e + N_t,
\end{equation}
where $B_t$ is the magnetic field produced by the targeted magnet at time $t$, $B_e$ is the uniform environmental field (e.g., Earth’s field and other persistent sources), and $N_t$ is high-frequency noise. 
The noise magnitude and frequency depend on nearby electronic devices but remain relatively stable, enabling effective suppression via filtering.

\subsubsection{Filtering $N_t$.}
\label{subsubsec:filtering}
To suppress the high-frequency noise component while maintaining real-time performance, {\name} adopts an exponential smoothing filter:
\begin{equation}
    \hat{Y}_{t+1} = \alpha Y_t + (1 - \alpha) \hat{Y}_t,
\end{equation}
where $Y_t$ is the observation at time $t$, $\hat{Y}_t$ is the estimate at time $t$, and $\alpha \in (0,1)$ is a smoothing parameter controlling the weight of past observations.  
In our experiments, $\alpha$ was empirically set to 0.5.  
This filter requires no future data, thereby ensuring low latency for real-time applications.


\subsubsection{Tackling the Earth's magnetic field $B_e$.}
Previous studies on mitigating environmental magnetic fields have mainly relied on physical modeling approaches, such as solving magnetic dipole equations~\cite{mag_dot,mag_x,magic_auto_calibration} or applying high-pass filters~\cite{auraring,finexus}.  
While dipole modeling can estimate magnet positions, it becomes inaccurate when the magnet deviates from the dipole assumption and typically requires $2n+1$ magnetometers to track $n$ magnets, leading to bulky layouts.  
To design a lightweight alternative, we address the problem of \emph{environmental bias} by making two practical assumptions: 
(1) the sensor orientation remains stable during the short duration of a self-touch, 
and (2) data collected before a magnet approaches the sensor captures the environmental magnetic field.  
These assumptions are supported by real-world observations: sensor orientation rarely changes within seconds, and the environmental field dominates when the target magnet is farther than about 15 cm from the sensor\cite{MagTouch}.

\begin{figure}[tp]
        \centering
        \includegraphics[width=\linewidth]{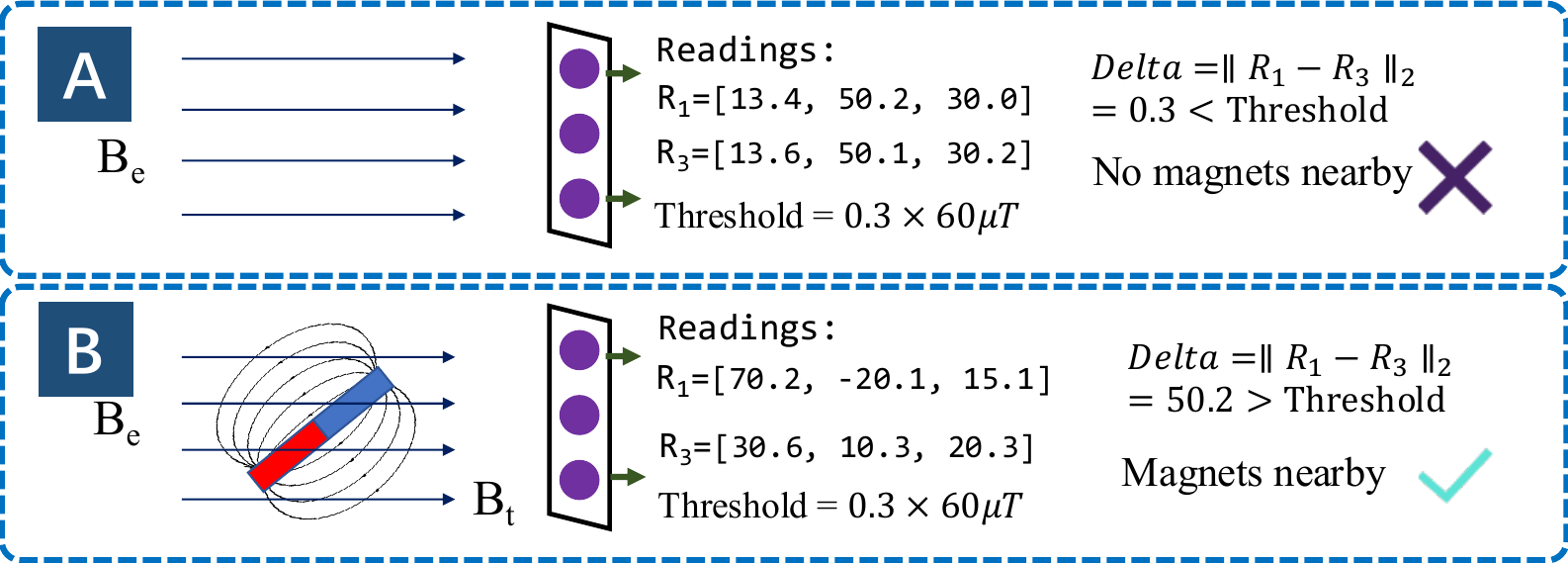}
        \vspace{0pt}
        \setlength{\abovecaptionskip}{-6pt}  
        \setlength{\belowcaptionskip}{-10pt}
        \caption{Schematic diagram of {\mdname}'s detection calculations in an environmental magnetic field with and without a targeted magnet.
        }
        \label{fig:mag_delta_diagram}
\end{figure}

Building on these assumptions, the next step is to detect when a target magnet enters the sensors' vicinity, so that the environmental field estimate can be separated from magnet-induced variations.  
To this end, we leverage discrepancies among multiple magnetometers: as illustrated in Fig.~\ref{fig:mag_delta_diagram}, readings remain consistent under a uniform field but diverge when a nearby magnet perturbs the field.  
Based on this principle, we design the following detection algorithm called {\mdname}.

\begin{algorithm}
\caption{\mdname: Detection of Nearby Magnets}
\begin{algorithmic}[1]
\State \textbf{Input:} Magnetometer readings $\{R_i, R_j\}$, threshold $T_{\Delta}$
\State \textbf{Output:} Presence of a nearby magnet (Boolean)
\For{each pair $(R_i, R_j)$ of magnetometer readings}
    \State $\Delta = \|R_i - R_j\|$
    \If{$\Delta > T_{\Delta}$}
        \State \textbf{Return:} Magnet detected
    \EndIf
\EndFor
\State \textbf{Return:} No magnet detected
\end{algorithmic}
\end{algorithm}


In {\mdname}, we maintain a 16-frame queue of sensor readings to estimate the background magnetic field.  
When the difference between any pair of magnetometers exceeds a threshold $T_{\Delta}$, the system detects the presence of nearby magnets.  
The oldest reading in the queue serves as an estimate of the environmental magnetic field, allowing {\name} to operate in real time.  

In practice, we empirically set $T_{\Delta}$ to $18\,\mu T$ based on preliminary trials.  
This value was sufficient to suppress false triggers caused by sensor noise while reliably detecting nearby magnets (e.g., up to 11 cm with magnetic rings).  
The threshold can be adapted to different environments or magnet types, and can be calibrated with a short initialization procedure of about five seconds of user movement.
Our empirical studies demonstrate that reliable detection of nearby magnets is critical for achieving high accuracy in self-touch sensing applications.






    

\subsection{Training Pipeline}
\input{figures/encoder_figures/encoder_figure}
To enhance adaptability to different users while minimizing labeled data requirements, we adopt a self-supervised pre-training and fine-tuning pipeline (Fig.~\ref{fig:finetune_chart}).  

\textbf{Pre-training.}  
We first collect unlabeled three-axis magnetometer readings from three sensors while ten users perform free-form movements. Each session lasts 45 seconds.  
The raw data is segmented into overlapping windows of length 16, resulting in input frames of size $(16, 9)$, where 9 corresponds to the three sensors on three axes.  
To preserve information about the absence of nearby magnets, we propose a magnetic-data-oriented normalization method: data are calibrated and normalized using the Earth’s magnetic field strength as the standard deviation, ensuring that zero readings remain unaffected.  
We employ the TS2Vec framework~\cite{yue2022ts2vec} with temporal and instance-wise contrastive loss~\cite{franceschi2019unsupervised,tonekaboni2021unsupervised,eldele2021time} to learn generalizable magnetic representations.
The encoder is trained using Adam~\cite{kingma2014adam} with a learning rate of 0.001 and converges within about 60 epochs, yielding robust features across users and gestures.

\textbf{Fine-tuning.}  
Once pre-trained, the encoder is transferred to the user’s device.  
For personalization, the user only needs to collect a few labeled samples (three per gesture, about 10 seconds of recording).  
These samples are used to fine-tune a lightweight downstream classifier such as an SVM.  
This design enables effective gesture recognition with minimal labeled data, substantially reducing the burden of per-user calibration.  

\textbf{Inference strategy.}  
As noted in Sec.~\ref{sec:pilot-study}, self-touches can vary significantly in duration.  
To handle this variability, we employ an event-based classification scheme.  
Specifically, when {\mdname} detects the approach of a nearby magnet, we continuously feed a 16-frame sliding window into the classifier.  
When the magnet departs, the predictions from all windows within the gesture are aggregated using a majority vote, and the final label is assigned.  
This strategy ensures robust recognition across gestures of different lengths while maintaining real-time responsiveness.


\section{Hardware Design of {\name}}
\label{sec:technical_hardware}

\begin{figure}[t]
    \centering
    \includegraphics[width=0.75\linewidth]{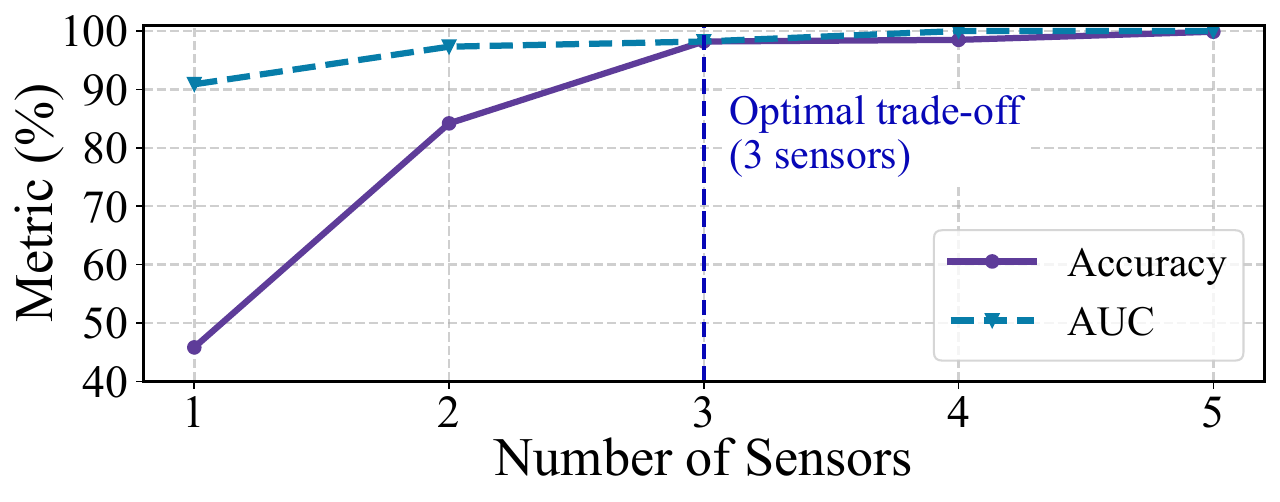}
    \setlength{\abovecaptionskip}{0pt}
    \setlength{\belowcaptionskip}{-20pt}
    \caption{Performance metrics of the classifier trained on data from eight simulated actions, using the macro average.}
    \label{fig:simul}
\end{figure}
We selected the MLX90393 MEMS magnetometer for its extensive measurement range (5-50 mT) and set the sample rate to 17 Hz, balancing temporal resolution with sensor noise. 
The microcontroller MDBT42Q collects data via SPI and transmits it to a computational unit, a Thinkpad X1 Yoga Gen 5 with an Intel i7-10510U CPU, via Bluetooth Low Energy (BLE). All models were computed on the CPU.
\begin{figure*}[th]
    \centering
    \setlength{\abovecaptionskip}{0pt}
    \includegraphics[width=0.92\linewidth]{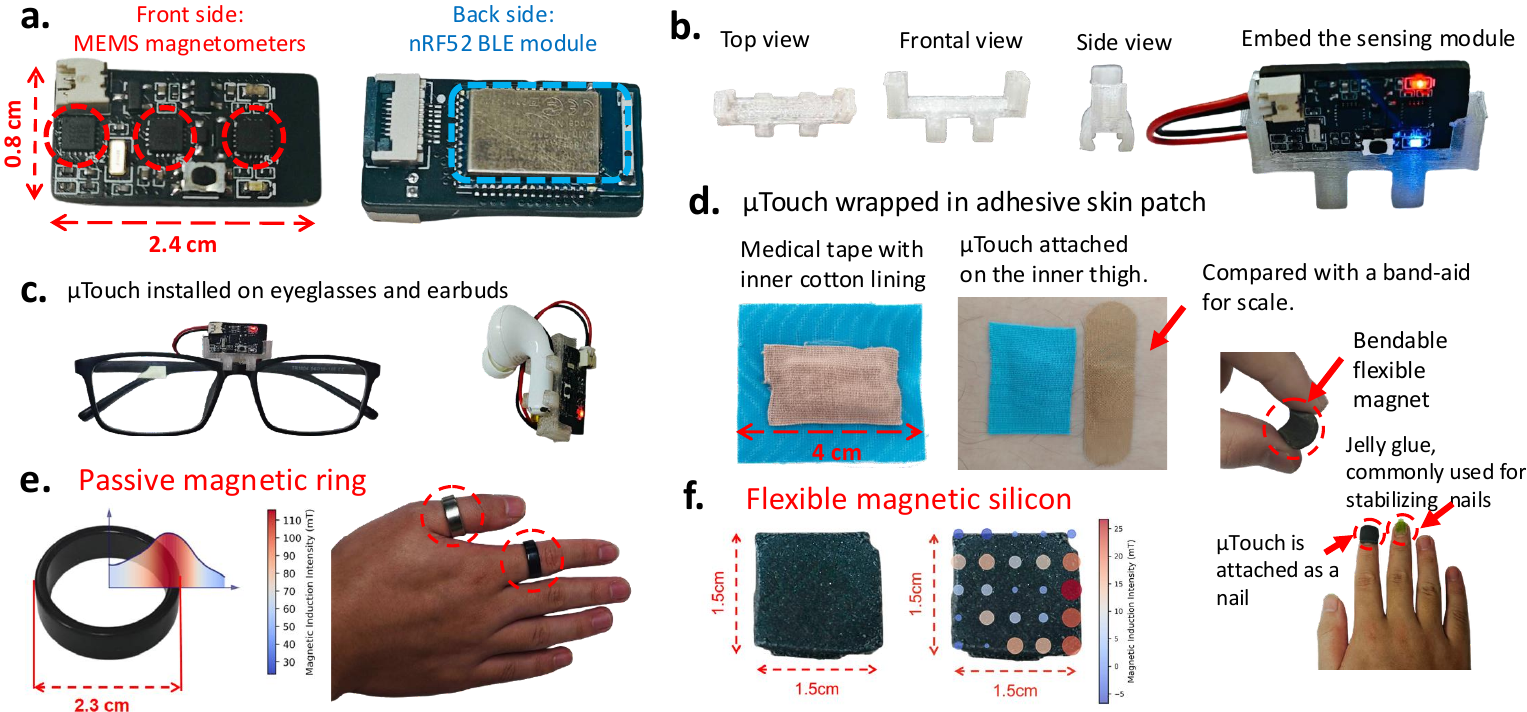}
    \caption{\textbf{Demonstration of {\name}'s hardware configuration.} \textbf{a,} Illustration of the sensing unit includes three MEMS magnetometers and a BLE module. 
    \textbf{b,} Different perspectives of our 3D-printed holder for embedding the sensing unit.
    \textbf{c,} Installing {\name} sensing unit on a regular pair of glasses.
    \textbf{d,} Integrating the module with a commodity-off-the-shelf wireless earbud.
    \textbf{e,} Illustration of the magnetic ring form factor and its magnetic field distribution.
    \textbf{f,} The flexible/bendable magnetic silicon form factor and its magnetic field distribution.
    }
    \label{fig:hardware_overview}
\end{figure*}

We conducted a \emph{simulation-guided computational design study} to explore the trade-off between sensor quantity and classification performance. 
In this framework, sensors were virtually placed along the x-axis at 0.8 cm intervals, and eight representative magnet movement directions were defined across a 360° plane (\{$0^\circ$, 45°, 90°, ...\}). 
Each trajectory started and ended 5 cm from the origin and was sampled 100 times with Gaussian noise (1 cm standard deviation) to emulate natural variations. 
A classifier was trained with 10 samples per action and evaluated on the remaining 90.  

As shown in Fig.~\ref{fig:simul}, classification accuracy rapidly increased with additional sensors, reaching 98.7\% with only three. 
Beyond this point, the performance gains exhibited \emph{diminishing returns}, indicating that further sensors would add hardware complexity without substantial accuracy benefits.  
This simulation-guided analysis informed our final choice of a three-sensor array, with 0.8 cm spacing, striking a balance between accuracy and form factor compactness.  
Next, we embedded the components on a four-layer printed circuit board (PCB), achieving a compact overall dimension of 2.4 cm by 1.2 cm (length by width). 
We utilized a dedicated, flexible printed circuit (FPC) connector for flashing purposes. 
The design is shown in Fig.~\ref{fig:hardware_overview}\textbf{a}.
This approach ensures {\name}'s lightweight form factor, enhancing its versatility for attachment to items like eyeglasses or adhesive skin patches, thereby improving wearability across different healthcare scenarios.

For applications that require mounting on wearable items such as eyeglasses, we designed a semi-enclosed 3D-printed casing to house the sensors.  
As shown in Fig.~\ref{fig:hardware_overview}\textbf{b}, the casing includes a pair of symmetrical circular grips that securely attach to the frame.  
In our tests, the grips remained sturdy even during active movements, preventing loosening or displacement.  

For the magnetic front-end design,
{\name} is compatible with a wide range of attachable passive magnets, e.g., magnetic rings on the fingers and magnetic silicon on the nail, as shown in Fig.~\ref{fig:hardware_overview}~(e, f). 
These passive magnets have two key features: (1) battery-free and (2) easily attachable.
We will elaborate on these design elements in Sec.~\ref{sec:applications}.


\section{Representative Applications}
\label{sec:applications}
To validate the efficacy of {\name}, we have selected two representative applications of self-touch detection: face-touching detection and body-scratch detection. 

\subsection{Fine-grained Face Touching Detection}
For fine-grained face-touching detection, we segmented the facial region into seven areas: the forehead, left eye, right eye, left cheek, right cheek, nose, and lips.
In addition, we explicitly introduced a \emph{no-touching} category to account for natural hand movements that should not be classified as face-touching, such as adjusting eyeglasses.  
By modeling this class during training, {\name} can suppress false positives and better reflect real-world usage scenarios.


We use magnetic rings as the magnetic front end for face-touching detection tasks.
These rings purchased online for approximately \$1 each, are axially magnetized.
As items that users can wear daily on their fingers, the rings effectively represent the hand movements of the user.  
Unlike dipole-model approaches that require $2n+1$ sensors to track $n$ magnets, {\name} achieves accurate classification with only three sensors.  
Using the {\mdname} algorithm (Sec.~\ref{subsec:tackling_the_environment_interference}), our system reliably detects magnetic rings within an effective range of approximately 11 cm.  
This allows practical deployment by positioning the sensing unit at the center of the eyeglass frame (Fig.~\ref{fig:hardware_overview}), ensuring coverage of all facial regions while maintaining a compact and unobtrusive design.

\subsection{Body-Scratch Detection}
\label{subsec:body_scratch}
\begin{figure}[t]
    \centering
\includegraphics[width=1\linewidth]{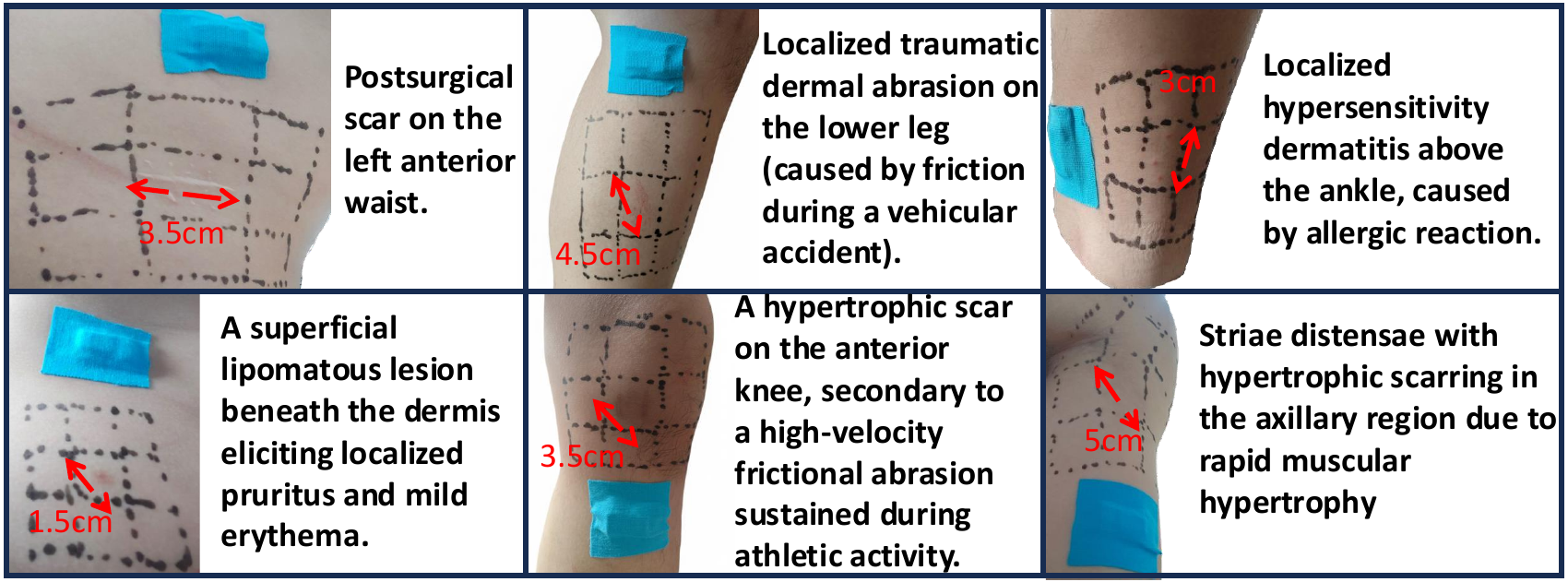}
\setlength{\abovecaptionskip}{-0pt}
    \caption{Illustration of the 3$\times$3 grid used for body-scratch detection. 
    The central cell corresponds to the affected skin area, while the surrounding cells represent non-affected regions.}
    \label{fig:body_scratch_grid}
\end{figure}

For body-scratch detection, we defined a 3$\times$3 grid on the participant’s skin near the affected region. 
The central cell corresponds to the patient’s dermatological lesion, which is frequently scratched. 
The surrounding cells represent non-affected regions and were included to model natural hand movements around, but not directly on, the lesion. 
This design allows {\name} to distinguish true scratching gestures from incidental nearby touches, thereby reducing false positives and improving detection reliability.

While magnetic rings are effective for face-touching detection, they are less suitable for body-scratch detection.  
Rings are positioned at the finger base, which provides stronger and more stable signals over a larger range --- useful for broad movements such as hand-to-face contact.  
Scratching gestures, however, are fine-grained and occur directly at the fingertip, where the distance from a ring would reduce sensitivity.  
To address this limitation, we introduce a lightweight, skin-conformal form factor attached to the fingernail, made of flexible magnetic silicon.  

The idea of creating flexible magnetic composites was introduced by ReSkin~\cite{bhirangi2022reskin}, which blended ferromagnetic powder with silicone rubber to produce a soft magnetic material.  
Inspired by this design, we adopt magnetic silicon as an alternative to rigid magnetic rings.  
Its fingertip placement and conformal form factor allow comfortable attachment near affected skin regions.  
Although its magnetic strength is weaker than that of rings, the closer and more natural positioning enables more fine-grained sensing of subtle scratching gestures, making it particularly suitable for healthcare scenarios such as body-scratch detection.  

We fabricated a $1.5\times1.5$ cm$^2$ piece of magnetic silicon.  
By attaching the sensor array to the skin with an adhesive patch (as shown in Fig.~\ref{fig:hardware_overview}\textbf{d}), {\name} enables reliable monitoring of scratching behaviors on or around dermatological lesion areas.  
This design makes it possible to detect subtle scratching gestures in daily life.

\begin{figure}[t]
\begin{minipage}{.99\linewidth}
    \centering
    
    \begin{minipage}{0.48\linewidth}
        \centering
        \vspace{-15pt}
        \includegraphics[width=\linewidth]{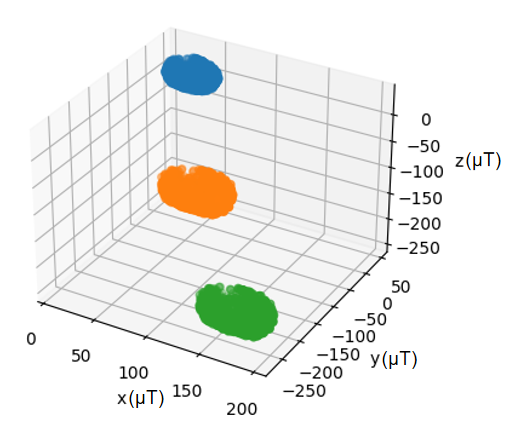}
        \vspace{-25pt}
        \caption*{(a) Before calibration.}
    \end{minipage}\hfill
    \begin{minipage}{0.48\linewidth}
        \centering
        \vspace{-15pt}
        \includegraphics[width=\linewidth]{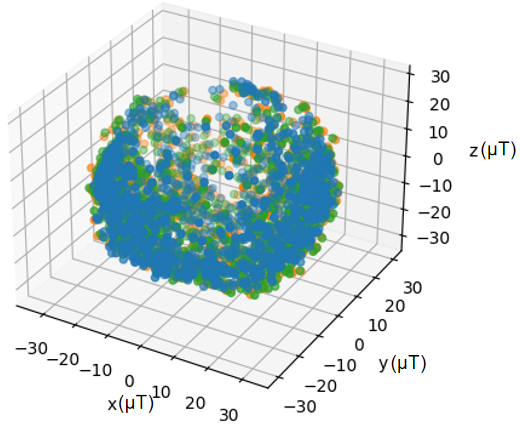}
        \vspace{-25pt}
        \caption*{(b) After calibration.}

    \end{minipage}
    \setlength{\abovecaptionskip}{0pt}
    \caption{Scatter plot of magnetometer readings before and after calibration in the presence of an external device with embedded magnets. 
In this example, the sensing unit was attached to AirPods Pro 2 while playing music.}
    \label{fig:calib_a_b}
\end{minipage}
\end{figure}

\subsection{Extensibility and Robustness}
Thanks to its compact hardware form factor and flexible machine-learning pipeline, {\name} can be readily adapted to diverse application scenarios.  
In our main applications, the sensing unit can be mounted on eyeglass frames for face-touching detection or directly on the skin with adhesive patches for body-scratch monitoring.  
Beyond these, {\name} can also be extended to other wearables, such as face masks, protective goggles, or even earbuds, depending on user needs.  
Users can further customize detection by fine-tuning the model with a few labeled samples.

In addition, {\name} is robust to interference from nearby electronic devices that contain embedded permanent magnets (e.g., headphones) or electromagnetic coils.  
While high-frequency noise can be suppressed via filtering (Sec.~\ref{subsubsec:filtering}), constant magnetic bias $B_{device}$ can be eliminated with a short one-time calibration.  
When the sensor array is fixed relative to the device, this bias remains stable and can be merged into the baseline offset.  
For example, when attached to AirPods Pro 2 \emph{during music playback}, the embedded magnets introduced a strong bias that undermined raw measurements.  
As shown in Fig.~\ref{fig:calib_a_b}, calibration recenters the sensor readings, effectively removing this device-induced bias.  
This procedure takes only about 10 seconds and needs to be performed once per device, making it lightweight and practical for everyday use.
\section{Evaluation}

\begin{figure}[t]
        \centering
        \includegraphics[width=0.95\linewidth]{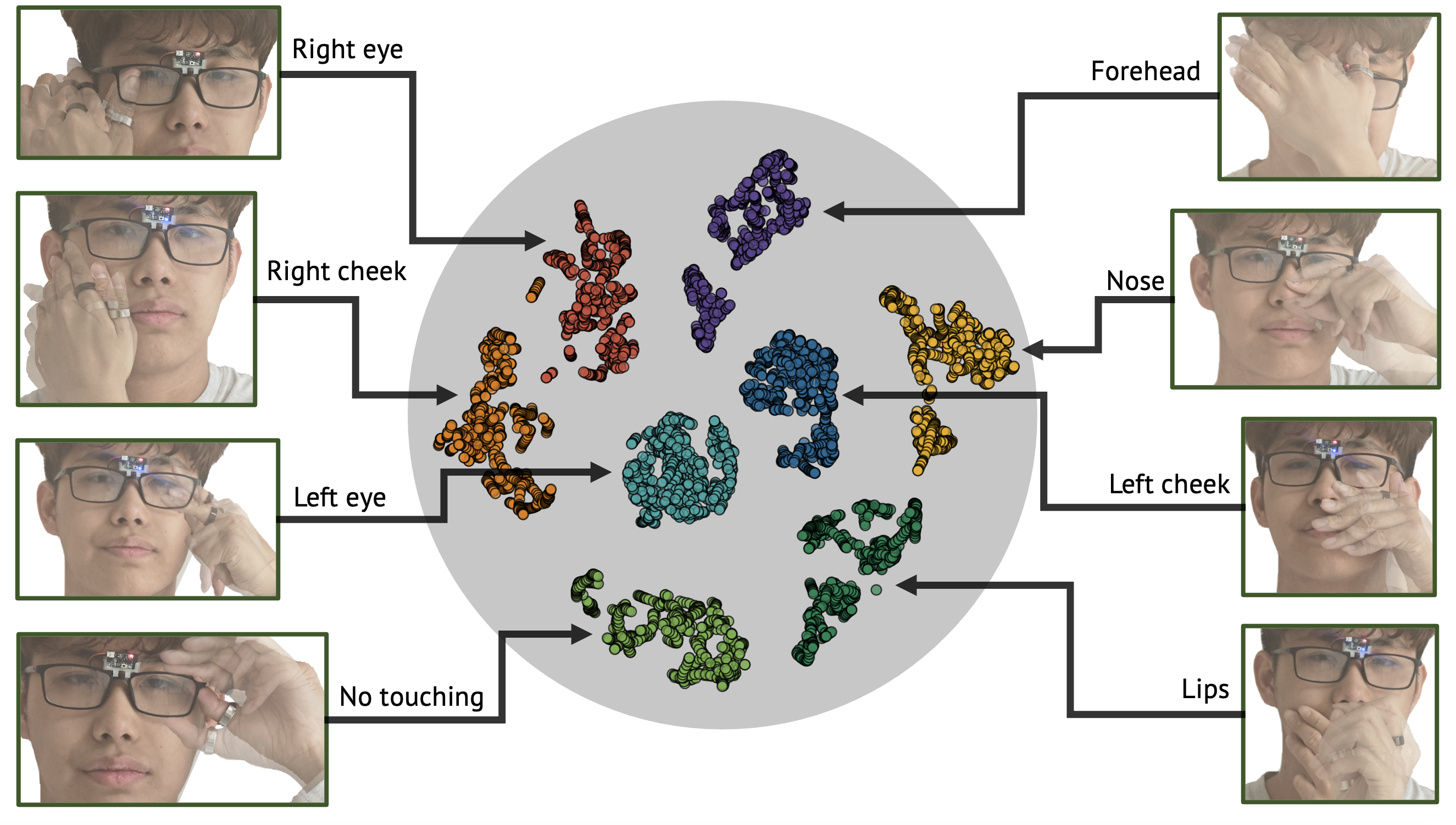}
        \setlength{\abovecaptionskip}{0pt}
        \caption{t-SNE visualization of face-touching gesture vectors from a pre-trained encoder.}
        \label{fig:t-sne_visualization}
\end{figure}
\label{sec:evaluation}

\label{sec:user_study}

We conducted two user studies to evaluate {\name} in healthcare-related self-touch detection tasks: \emph{fine-grained face-touching detection} and \emph{body-scratch detection}. 
All tests in this work were approved by the institutional IRB Board.

\subsection{Participants}
For the face-touching study, we recruited 11 student volunteers aged 19–24.  
All participants were right-handed; to assess both dominant and non-dominant hand usage, five wore magnetic rings on the left hand and six on the right hand.  

For the body-scratch study, we recruited 12 student volunteers aged 19–26.  
Six participants reported dermatological conditions (e.g., eczema or skin irritation) or a medical history that led to habitual scratching of specific body regions, as shown in Fig.~\ref{fig:body_scratch_grid}.  
The other six participants did not report scratching problems; instead, each was paired with one symptomatic participant and instructed to mimic the same scratching actions on the corresponding body region.  
This paired design (1) establishes a control group to prevent {\name} from overfitting to participant-specific behaviors, and (2) expands scratching samples without repeatedly aggravating the condition of symptomatic participants, ensuring both ethical safety and data diversity.  
To further account for hand dominance, every participant performed tasks with both the dominant and non-dominant hand.  
For scratching actions that could not be naturally swapped across hands (e.g., scratching the left forearm with the left hand), we mirrored the target body region so that equivalent actions could still be performed by the opposite hand.

\subsection{Data Collection Protocol}
In both studies, data collection followed a three-phase procedure: 

\noindent\textbf{(1) Pre-training:}  
We first pre-trained a magnetic motion encoder using 45-second segments of arbitrary hand or touch-like movements, which capture natural magnetic signal patterns.  
To validate generalization, the encoder was always trained without the target participant’s data.  
For the face-touching study, pre-training used data from 10 participants, excluding the test subject.  
For the body-scratch study, we did not collect additional pre-training data; instead, we directly reused the encoder pre-trained on the face-touching dataset (11 participants).  
This setup allowed us to evaluate the encoder under both cross-user and cross-task conditions, demonstrating robustness across participants and applications.

\noindent\textbf{(2) Training:}  
Participants then performed target gestures within a three-second window.  
For face-touching, this included seven facial regions (forehead, eyes, cheeks, nose, lips), plus a ``no touching'' action (adjusting glasses).  

For body-scratch detection, we considered two complementary classification settings:  
(1) a \emph{binary setting} (\textit{scratch} vs. \textit{no scratch}), which directly reflects the clinical need of monitoring scratching behaviors; and  
(2) a \emph{fine-grained nine-class setting} based on the 3$\times$3 grid, where the central cell corresponded to the lesion and the surrounding cells represented non-affected regions.  
The binary setting evaluates whether {\name} can reliably detect scratching versus incidental touches, while the nine-class setting assesses its ability to distinguish scratching of the lesion from nearby, non-lesion touches, thereby reducing false positives and improving robustness.  

In the binary setting, participants performed three scratching actions on the lesion area and one ``no scratch'' action on a randomly chosen non-affected grid cell.  
In the fine-grained setting, participants performed scratching actions on each of the nine grid cells.  
Each gesture lasted about three seconds.  

For both face-touching and body-scratch tasks, models were first trained using data from other participants to validate cross-user generalization, and then fine-tuned with less than 30 seconds of each participant’s own data.

\noindent\textbf{(3) Testing:} Each participant completed five sessions.  
In each session, they performed all gesture categories once (face: seven regions + no touching; body: scratching + no scratching).  
To assess robustness under varying environmental magnetic fields, each session was conducted in one of five evenly spaced orientations around 360°.  
Gesture durations were not fixed, allowing natural variation, i.e., face: 3 - 15s; body: 2 - 12s.

\begin{table*}[t]
  \small
  \centering
\caption{Results of different models on face-touching Detection and body scratch Detection. 
All numbers are percentages (\%). 
``Bin.Acc.'' denotes binary classification accuracy (scratch vs. no-scratch) for the body-scratch task.}
  \label{tab:main_results1}
  \setlength{\tabcolsep}{5pt}
{%
\begin{tabular}{@{\hskip0pt}l@{\hskip4pt}g@{\hskip2pt}c@{\hskip2pt}c@{\hskip4pt}c@{\hskip4pt}c@{\hskip4pt}o@{\hskip4pt}g@{\hskip4pt}c@{\hskip4pt}c@{\hskip4pt}c@{\hskip4pt}c@{\hskip0pt}}
    \toprule
    \multirow{2}{*}{Models} 
      & \multicolumn{5}{c}{Face Touching Detection} 
      & \multicolumn{6}{c}{Body Scratch Detection} \\
    \cmidrule(lr){2-6} \cmidrule(lr){7-12}
      & Accuracy & ~F1 Score~ & Precision & Recall & AUC 
      & Bin.Acc. & Accuracy & ~F1 Score~ & Precision & Recall & AUC \\
    \midrule
    Encoder+SVM     & \textbf{93.41} & \textbf{93.39} & \textbf{93.47} & \textbf{93.41} & \textbf{98.77} & \textbf{94.58} & \textbf{94.63} & \textbf{94.71} & \textbf{95.06} & \textbf{94.63} & \textbf{99.92} \\
    SVM             & 90.71 & 90.83 & 91.38 & 90.71 & 98.24 & 83.33 & 83.33 & 83.79 & 85.63 & 83.33 & 97.24 \\
    Random Forest   & 86.82 & 86.91 & 87.27 & 86.82 & 98.73 & 89.17 & 87.96 & 88.23 & 89.42 & 87.96 & 97.91 \\
    PCA+SVM         & 91.14 & 91.15 & 91.54 & 91.14 & 98.50 & 85.00 & 85.19 & 85.63 & 87.11 & 85.19 & 98.47 \\
    PCA+RF          & 82.38 & 88.78 & 89.10 & 89.00 & 98.20 & 79.17 & 79.44 & 80.07 & 81.89 & 79.44 & 98.13 \\
    \midrule
    Non-Dom hand    & 89.50 & 89.34 & 89.64 & 89.50 & 98.10 & 95.00 & 95.19 & 95.25 & 95.68 & 95.19 & 99.94 \\
    Dom hand        & 96.67 & 96.68 & 96.82 & 96.67 & 99.30 & 94.17 & 94.07 & 94.13 & 94.58 & 94.07 & 99.91 \\
    \midrule
    w/o MagDelta    & 88.18 & 88.10 & 88.58 & 88.18 & 98.25 & 45.00 &  49.26 & 49.71 & 51.73 & 49.26 & 89.65 \\
    \textit{$\Delta$} & \textcolor[HTML]{5E3C99}{(-5.23)} 
                      & \textcolor[HTML]{5E3C99}{(-5.29)} 
                      & \textcolor[HTML]{5E3C99}{(-4.89)} 
                      & \textcolor[HTML]{5E3C99}{(-5.23)} 
                      & \textcolor[HTML]{5E3C99}{(-0.52)}
                      & \textcolor[HTML]{5E3C99}{(-49.58)} 
                      & \textcolor[HTML]{5E3C99}{(-45.37)} 
                      & \textcolor[HTML]{5E3C99}{(-45.00)} 
                      & \textcolor[HTML]{5E3C99}{(-43.33)} 
                      & \textcolor[HTML]{5E3C99}{(-45.37)} 
                      & \textcolor[HTML]{5E3C99}{(-10.27)} \\
    \midrule
    Before Remount  & 97.50 & 97.51 & 97.57 & 97.50 & 99.44 & 94.17 & 95.19 & 95.25 & 95.68 & 95.19 & 99.94 \\
    After Remount   & 97.14 & 97.14 & 97.29 & 97.14 & 99.70 & 93.33 & 94.81 & 94.88 & 95.31 & 94.81 & 99.78 \\
    \textit{$\Delta$} & \textcolor[HTML]{5E3C99}{(-0.36)} 
                      & \textcolor[HTML]{5E3C99}{(-0.37)} 
                      & \textcolor[HTML]{5E3C99}{(-0.28)} 
                      & \textcolor[HTML]{5E3C99}{(-0.36)} 
                      & \textcolor[HTML]{077DA9}{(+0.26)}
                      & \textcolor[HTML]{5E3C99}{(-0.84)}
                      & \textcolor[HTML]{5E3C99}{(-0.38)} 
                      & \textcolor[HTML]{5E3C99}{(-0.37)} 
                      & \textcolor[HTML]{5E3C99}{(-0.37)} 
                      & \textcolor[HTML]{5E3C99}{(-0.38)} 
                      & \textcolor[HTML]{5E3C99}{(-0.16)} \\
    \midrule
    Before Follow up & 89.00 & 88.78 & 89.10 & 89.00 & 98.07 & 94.17 & 95.19 & 95.25 & 95.68 & 95.19 & 99.94 \\
    Follow up Study  & 90.50 & 90.47 & 90.86 & 90.50 & 98.20 & 93.33 & 92.59 & 92.63 & 92.86 & 92.59 & 99.77 \\
    \textit{$\Delta$}  & \textcolor[HTML]{077DA9}{(+1.50)} 
                       & \textcolor[HTML]{077DA9}{(+1.69)} 
                       & \textcolor[HTML]{077DA9}{(+1.76)} 
                       & \textcolor[HTML]{077DA9}{(+1.50)} 
                       & \textcolor[HTML]{077DA9}{(+0.13)}
                       & \textcolor[HTML]{5E3C99}{(-0.84)} 
                       & \textcolor[HTML]{5E3C99}{(-2.60)} 
                       & \textcolor[HTML]{5E3C99}{(-2.62)} 
                       & \textcolor[HTML]{5E3C99}{(-2.82)} 
                       & \textcolor[HTML]{5E3C99}{(-2.60)} 
                       & \textcolor[HTML]{5E3C99}{(-0.17)} \\
    \bottomrule
\end{tabular}%
}
\vspace{-2ex}
\end{table*}

\subsection{Classification Performance}
\label{sec:results_baselines}

\begin{figure}[t]

    \setlength{\belowcaptionskip}{-12pt}
    \centering
    \includegraphics[width=1\linewidth]{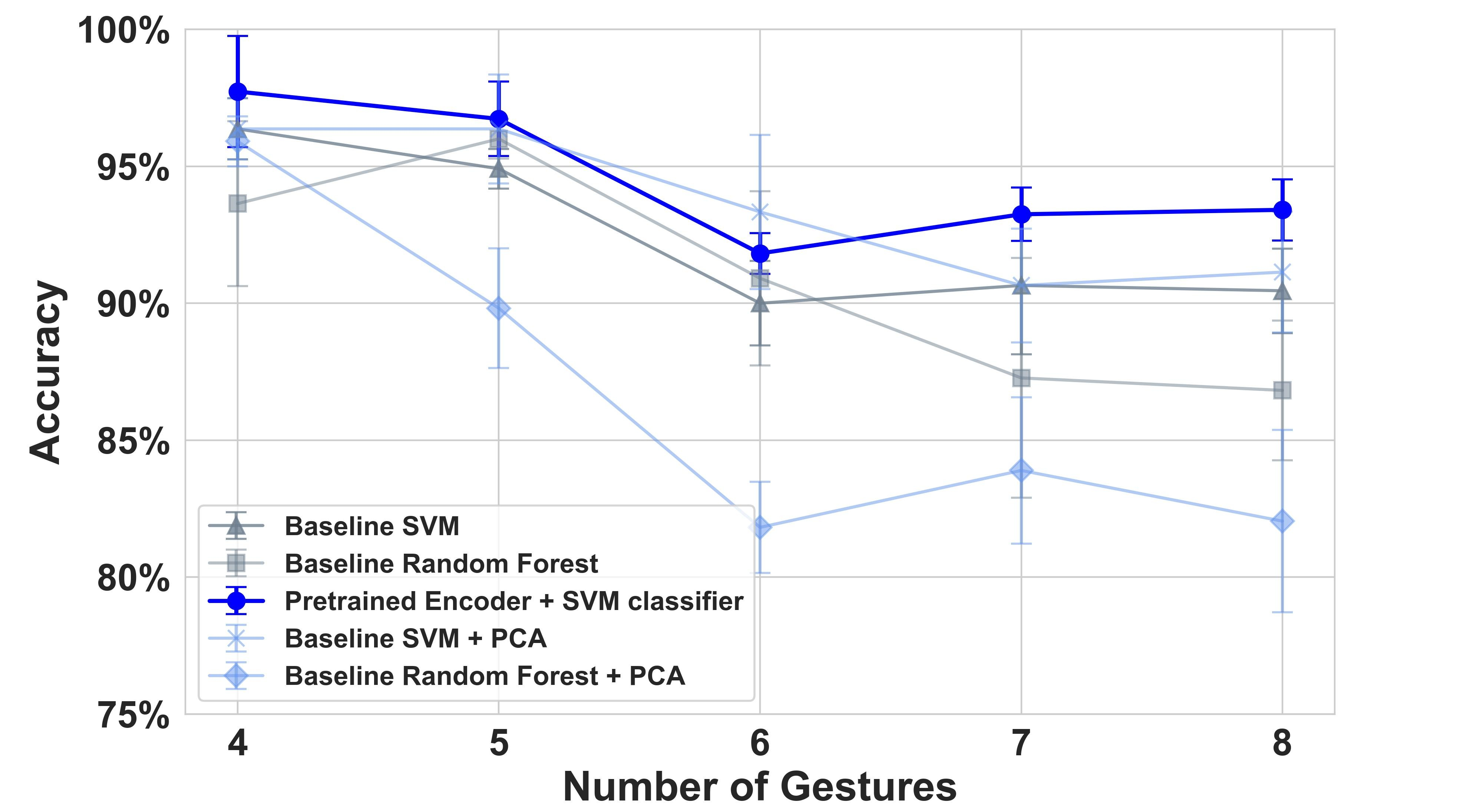}
    \caption{
    {\name}'s overall accuracy with the varying number of gestures. Overall accuracy is calculated from all gestures in five different orientations for all volunteers.}
    \label{fig:face_touching_baseline}
\end{figure}

We compared {\name} against conventional classifiers, including SVM, Random Forest, and PCA-based variants (Table~\ref{tab:main_results1}).  
Across both tasks, {\name} consistently achieved the highest accuracy.  

For \textbf{face-touching detection}, {\name} reached 93.4\% accuracy, outperforming baselines by 2–6\%.  
As Fig.~\ref{fig:face_touching_baseline} shows, even when the number of gesture classes increased to eight (seven facial regions plus ``no touching''), accuracy dropped by only 4.1\%, indicating strong robustness.  
The most challenging case was the ``lips'' region, which showed higher confusion with ``nose'' due to weaker signal strength.  

For \textbf{body-scratch detection}, {\name} achieved 94.6\% accuracy, while baselines generally remained below 88\%.  
Both the binary (scratch vs. no scratch) and fine-grained grid settings maintained high accuracy, demonstrating the system’s ability to separate true scratching from incidental nearby touches.  
In particular, {\name} reached a binary accuracy of 94.6\%, exceeding baseline models by more than 5–15\%, which highlights its reliability in healthcare scenarios where binary detection is most critical.

Overall, {\name} improved accuracy by 5–10\% over traditional baselines and scaled reliably across different tasks and gesture complexities.

To further understand why the encoder+SVM pipeline outperforms traditional baselines, we visualize the learned feature representations using t-SNE (Fig.~\ref{fig:t-sne_visualization}).  
Each cluster corresponds to one of the eight face-touching categories.  
The visualization shows that the pre-trained encoder maps gestures into compact and well-separated clusters, enabling simple classifiers such as SVM to achieve high accuracy.  
In contrast, traditional approaches like PCA or Random Forest rely on raw or hand-crafted features, which produce less separable distributions and thus lower classification performance.  
Notably, even visually similar gestures (e.g., ``nose'' vs. ``lips'') form distinguishable clusters, highlighting the encoder’s ability to capture fine-grained differences in magnetic field dynamics.  
This interpretability further validates the robustness of {\name}'s representation learning approach.

\subsubsection{Dominant vs. Non-dominant Hands}
We analyzed the effect of using dominant versus non-dominant hands.

For the \textbf{face-touching detection} task, participants using their dominant hand achieved higher accuracy (96.7\%) than those using non-dominant hand (89.5\%).  
This discrepancy likely arises because non-dominant hand movements were less consistent with training data, reflecting limited dexterity in replicating subtle face-touching gestures.
Nevertheless, all non-dominant results still exceeded 89\%, confirming that {\name} remains reliable in this setting.  

For \textbf{body-scratch detection}, we designed a mirrored protocol to accommodate regions that cannot be scratched symmetrically (e.g., scratching the left arm with the left hand).  
Here, the impact of hand dominance was less pronounced: both dominant and non-dominant hand conditions maintained accuracy above 94\%.

\subsubsection{Ablation Study}
\label{subsubsec:facetouching_ablation}
Mitigating environmental magnetic fields, particularly the Earth’s magnetic field, is essential for robust performance under varying orientations.  
To evaluate the importance of our environmental field mitigation algorithm {\mdname} (Sec.~\ref{subsec:tackling_the_environment_interference}), we conducted an ablation study by directly feeding raw sensor readings to the classifier.  

As shown in Table~\ref{tab:main_results1}, removing {\mdname} led to a moderate accuracy drop in \textbf{face-touching} (from 93.41\% to 88.18\%).  
Here, the stronger magnetic field generated by the magnetic rings still provided separable signals, even though robustness across orientations degraded.  

In contrast, the effect was dramatic in the \textbf{body-scratch task}, where accuracy plummeted from 94.63\% to 49.26\%.  
We attribute this sharp decline to two factors: (1) the use of magnetic fingernails made of flexible silicon, which produce significantly weaker magnetic fields than rings, and (2) the flexible sensor placement on different body regions, which increases exposure to environmental bias.  
Without {\mdname}, the weak target signals were easily overwhelmed by the Earth’s field and background interference.  

These results demonstrate that while {\mdname} enhances robustness in all cases, it is indispensable when weak magnets and diverse sensor placements are involved.  
This finding underscores the necessity of lightweight environmental compensation algorithms for healthcare applications such as body-scratch detection.

\subsubsection{Remount Experiment}
\label{sec:remount}
To evaluate robustness in real-world usage, we conducted a remount experiment across both tasks.  
After completing the initial trials, participants were asked to remove {\name} and the magnets, rest for at least five minutes, and then reattach the device without assistance from the research team.  
This procedure emulates daily usage, where users may need to reposition or reapply the system themselves.  

In the \textbf{face-touching study}, seven participants (five dom-hand, two non-dom-hand) took part, and {\name} maintained stable performance with accuracy changing only slightly from 97.50\% before remount to 97.14\% after remount.  
In the \textbf{body-scratch study}, all six participants with dermatological conditions performed the remount procedure on their affected regions.  
Despite variability in sensor placement and lesion location, accuracy remained high (94.88\%), confirming that {\name} sustains reliable performance after reattachment.

These results demonstrate that {\name} is resilient to minor placement variations and remains reliable even when reattached by users, including patients with real dermatological conditions, without expert calibration.

\subsubsection{One-month Follow-up Study}
\label{subsubsec:followup}

\begin{figure}[t]
    \centering
    \includegraphics[width=0.98\linewidth]{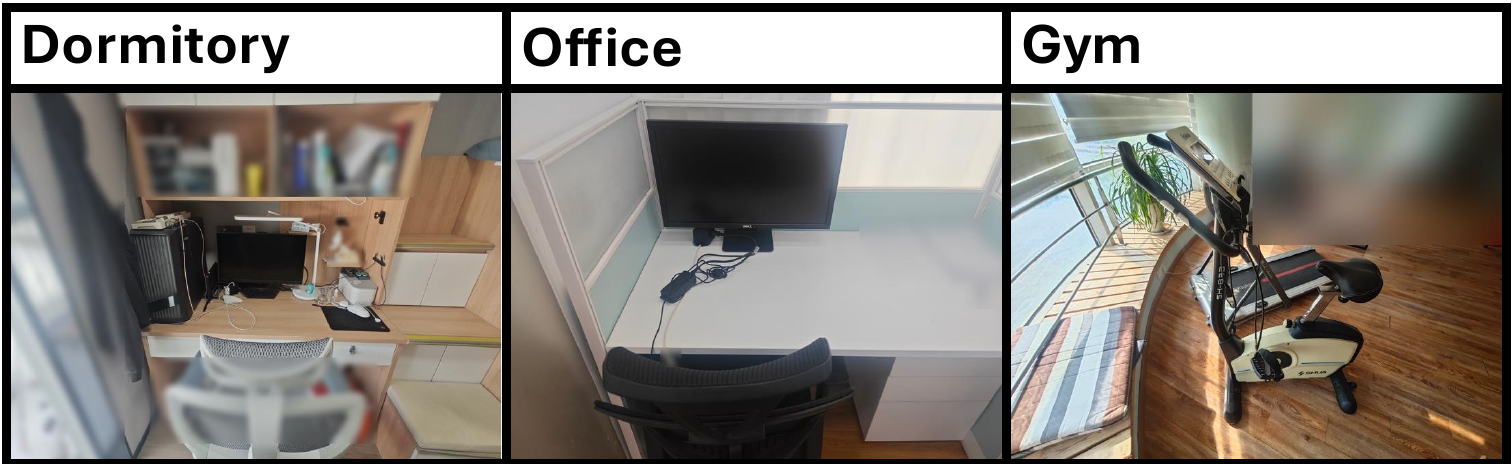}
    \caption{Three environments where {\name} was tested: 
    (1) Dormitory, with hardware interference from everyday electronics; 
    (2) Office; 
    (3) Gym, with fitness equipment built from ferromagnetic materials and highly mobile usage.}
    \label{fig:environment}
\end{figure}

To assess long-term robustness in realistic conditions, we conducted a
follow-up study without collecting new training data.
For face-touching, five participants repeated the task one month later in the gym
environment (Fig.~\ref{fig:environment}), where accuracy remained stable despite stronger interference
(89.0\% initially vs.\ 90.5\% in the follow-up).
For body-scratch detection, six participants with dermatological conditions repeated
the task one month later in office and dormitory environments, and {\name}
maintained accuracy above 92\% despite variations in sensor placement and
background interference.

\textbf{In-the-wild evaluation.}
To further evaluate naturalistic usage beyond scripted runs,
we conducted an in-the-wild face-touch study with three participants
from the original cohort in two environments (office and gym).
Each participant wore {\name} for approximately one hour per environment,
behaving naturally (e.g., working or exercising) without scripted actions.
Using the same per-participant models,
{\name} achieved a precision above 92\%.

These results highlight {\name}’s ability to sustain performance across time, environments, and tasks.  
The stability of accuracy without retraining demonstrates that the combination of a pre-trained encoder with a lightweight SVM classifier provides a strong balance of generalization and adaptability, making {\name} well-suited for real-world deployment.

\subsection{Energy Efficiency and Latency}
\label{sec:energy_efficiency}
We measured the power consumption of {\name} by monitoring the voltage and current drawn from its battery under continuous operation.  
The whole sensing hardware, which includes three magnetometers, the control circuit, and the BLE module, consumed on average 25\,mW.  
With a compact 3.7\,V, 100\,mAh battery, this corresponds to approximately 8 hours of uninterrupted operation, demonstrating suitability for daily use.  

To evaluate runtime performance, we benchmarked the end-to-end ML inference pipeline on a Xiaomi 13 smartphone with a Qualcomm SM8550-AB chipset.  
Inference was executed in a streaming setting, where 1000 test samples were processed sequentially.  
The total runtime was 53.37 seconds, corresponding to an average latency of 53.37\,ms per sample.  
Considering the system’s sampling rate of 17\,Hz (58.8\,ms period), this latency introduces only marginal overhead, confirming that {\name} can operate in real time on commodity mobile hardware.


\vspace{-1pt}

\section{Related Works}
\label{sec:related_works}

{\name} implements passive magnetic sensing for detecting micro gestures. Related works include face-touching detection, one-handed gesture interaction, magnetic sensing. 

\subsection{Self-touch Detection}
Prior work on self-touch detection has explored diverse sensing modalities for detecting face-touching~\cite{DAurizio2020Preventing,ActSonic}.
COVID-away~\cite{covid-away} employs a multi-sensor wearable extracting over 100 features from motion and pressure signals, but only distinguishes touching from non-touching.  
SavingFace~\cite{rojas2021scalable} leverages acoustic sensing via wired earbuds but is limited to coarse detection.  
FaceSense~\cite{face_sense} combines thermal imaging with physiological signals (impedance, EMG), yet mainly addresses binary classification and requires bulky hardware.  
D-Touch~\cite{D_touch} achieves finer granularity with a neck-mounted infrared camera, recognizing up to 17 facial activities, but at the cost of high power consumption and privacy concerns.  

These approaches highlight trade-offs of existing modalities: while some achieve fine-grained classification, they rely on intrusive, power-hungry, or privacy-sensitive sensors; others remain limited to binary detection.

\subsection{Magnetic Sensing}
\noindent\textbf{Passive Magnet Tracking}. MagX~\cite{mag_x} determines magnet positions using passive magnets and 8 magnetometers, achieving 0.76 cm positional error for tracking one magnet within the range of 11 cm. MagDot~\cite{mag_dot}, with arrow-shaped magnetometer arrays, tracks joint angles with a mean error below $1^\circ$. Both employ the Levenberg-Marquardt (LM) algorithm~\cite{lm_gradient_descent} for dipole modeling.

\noindent\textbf{Electromagnet Tracking}. Electromagnetic tracking offers high accuracy, as alternating magnetic fields are stable and resistant to environmental noise. Finexus~\cite{finexus} leverages multiple electromagnetic frequencies and filters to track finger movements, while AuraRing~\cite{auraring} uses a coil in a ring and magnet sensors in a wristband to achieve 3-DoF position and 2-DoF orientation tracking with iterative models anda  neural network.
However, electromagnetic tracking has notable drawbacks compared to {\name}'s battery-free solution. 
It requires a powered coil on the finger, thus extensively increasing energy consumption and generating substantial heat, with coils potentially reaching temperatures high enough to melt 3D-printed PLA enclosures (melting point around 180°C).
The receiver is usually bulkier compared with {\name}. For example, AuraRing~\cite{auraring} needs a wrist-sized receiver.

\section{Conclusion}
\label{sec:conclusion}
This paper presents {\name} for sensing self-touch gestures.
{\name} allows users to quickly expand their designated gesture set using a novel self-supervised learning approach.
In our exemplary applications, {\name} showcases its adaptability to various form factors, e.g., eyeglasses and skin patches, in two representative usage scenarios.
We believe {\name} demonstrates a feasible path towards practical self-touch gesture sensing in real-world applications.

\section*{Acknowledgment}
This work was supported by the National Natural Science Foundation of China under Grants No. 62472283. This work was also supported by the Cornell China Center’s joint seed grant.

\balance{}

\bibliographystyle{IEEEtran}
\bibliography{ref}

\end{document}